\documentclass[journal]{IEEEtran}

\usepackage{cite}
\usepackage{amsmath,amssymb,amsfonts,bm}
\usepackage{algorithm}
\usepackage{algorithmic}
\usepackage{graphicx}
\usepackage{epstopdf}
\usepackage{booktabs}
\usepackage{makecell}
\usepackage{multirow}
\usepackage{textcomp}
\usepackage{xcolor}
\usepackage[colorlinks  = true,
            linkcolor   = black,
            urlcolor    = magenta,
            anchorcolor = blue,
            bookmarks   = false
            ]{hyperref}

\def\BibTeX{{\rm B\kern-.05em{\sc i\kern-.025em b}\kern-.08em
    T\kern-.1667em\lower.7ex\hbox{E}\kern-.125emX}}

\newcommand{\overbar}[1]{\mkern 1.5mu\overline{\mkern-1.5mu#1\mkern-1.5mu}\mkern 1.5mu}

\begin{document}

\title{Deep Learning for Hybrid Beamforming with Finite Feedback in GSM Aided mmWave MIMO Systems}

\author{

Zhilin Lu, Xudong Zhang, Rui Zeng and Jintao Wang,~\IEEEmembership{Senior Member,~IEEE}

\thanks{

The authors are with the Department of Electronic Engineering, Tsinghua University, and Beijing National Research Center for Information Science and Technology (BNRist), Beijing 100084, China. (e-mail: luzl18@mails.tsinghua.edu.cn, zxd22@mails.tsinghua.edu.cn, zengr21@mails.tsinghua.edu.cn, wangjintao@tsinghua.edu.cn).

The key results can be reproduced with the following github link: \textnormal{\href{https://github.com/Kylin9511/GsmEFBNet}{https://github.com/Kylin9511/GsmEFBNet}}.
}
}

{}

\maketitle

\begin{abstract}
Hybrid beamforming is widely recognized as an important technique for millimeter wave (mmWave) multiple input multiple output (MIMO) systems. Generalized spatial modulation (GSM) is further introduced to improve the spectrum efficiency. However, most of the existing works on beamforming assume the perfect channel state information (CSI), which is unrealistic in practical systems. In this paper, joint optimization of downlink pilot training, channel estimation, CSI feedback, and hybrid beamforming is considered in GSM aided frequency division duplexing (FDD) mmWave MIMO systems. With the help of deep learning, the GSM hybrid beamformers are designed via unsupervised learning in an end-to-end way. Experiments show that the proposed multi-resolution network named GsmEFBNet can reach a better achievable rate with fewer feedback bits compared with the conventional algorithm.
\end{abstract}


\section{Introduction}

\IEEEPARstart{T}{he} millimeter wave (mmWave) multiple-input multiple-output (MIMO) technique is of great significance for the 5$^\text{th}$ generation (5G) wireless communication system to increase the available bandwidth \cite{mansoor20175g}. With a limited number of radio frequency (RF) chains, hybrid beamforming is necessary to compensate for the free-space pathloss in mmWave MIMO systems. Meanwhile, generalized spatial modulation (GSM) is introduced to MIMO systems with only part of the transmitting antennas activated \cite{he2015infty}. The spectral efficiency is improved under the GSM scheme since extra information can be carried by the choice of the activated antennas.

Further, the beamforming of GSM aided MIMO systems arouses great attention. Hybrid beamformers are designed with the gradient ascent algorithm in \cite{he2018spatial} to optimize the sum rate of GSM aided mmWave MIMO systems. A novel algorithm for GSM hybrid beamforming is introduced in \cite{lu2018low} based on turbo optimization. However, most of these existing works assume that perfect channel state information (CSI) is known, which is not practical in real systems.

On the other hand, deep learning (DL) is widely applied to the beamforming task and achieves compelling superiority. For instance, hybrid precoders and combiners are learned based on the convolutional neural network (CNN) for mmWave MIMO systems in \cite{elbir2019cnn}. A two-stage hybrid beamforming algorithm is designed with the help of CNN and outperforms a series of traditional benchmarks in \cite{liu2021two}.

It is worth noting that the practical downlink CSI acquisition is included in some recent works on DL aided beamforming \cite{guo2021deep-FB-BF,Sohrabi2021deep,Gao2022data, lu2023towards}. These works prove that the channel \textit{Estimation}, \textit{Feedback}, and \textit{Beamforming} (EFB) can be optimized as a whole via end-to-end training. Yet simple digital beamformer is used in \cite{Sohrabi2021deep} while classic hybrid beamformers are considered in \cite{Gao2022data,lu2023towards}. In other words, the research on DL-based GSM beamforming with downlink CSI obtainment is still lacking.

In this letter, the DL-based EFB joint optimization under the GSM scheme is looked into for frequency division duplexing (FDD) mmWave MIMO systems. Notably, the CSI compressed feedback is necessary since the uplink and downlink channels are asymmetric for the FDD modes. The main contributions of the paper are summarized below.

\begin{itemize}
  \item Practical downlink CSI acquisition is first considered for the GSM aided hybrid beamforming task under the DL-based EFB pipeline with finite feedback capacity.
  \item A multi-resolution network named GsmEFBNet is specially designed with adaptive kernel size and outperforms the traditional benchmark under various feedback bits and different signal-to-noise ratios (SNR).
\end{itemize}

The rest of the paper is arranged as follows. Section \ref{Section-SystemModel} explains the system model and section \ref{Section-Algorithm} introduces the proposed GsmEFBNet for the EFB joint optimization pipeline in detail. The numerical results are analyzed in section \ref{Section-Experiment} while the whole paper is concluded in section \ref{Section-Conclusion}.

\section{System Model} \label{Section-SystemModel}

\begin{figure}[!t]
  \centering
  \includegraphics[width=\linewidth]{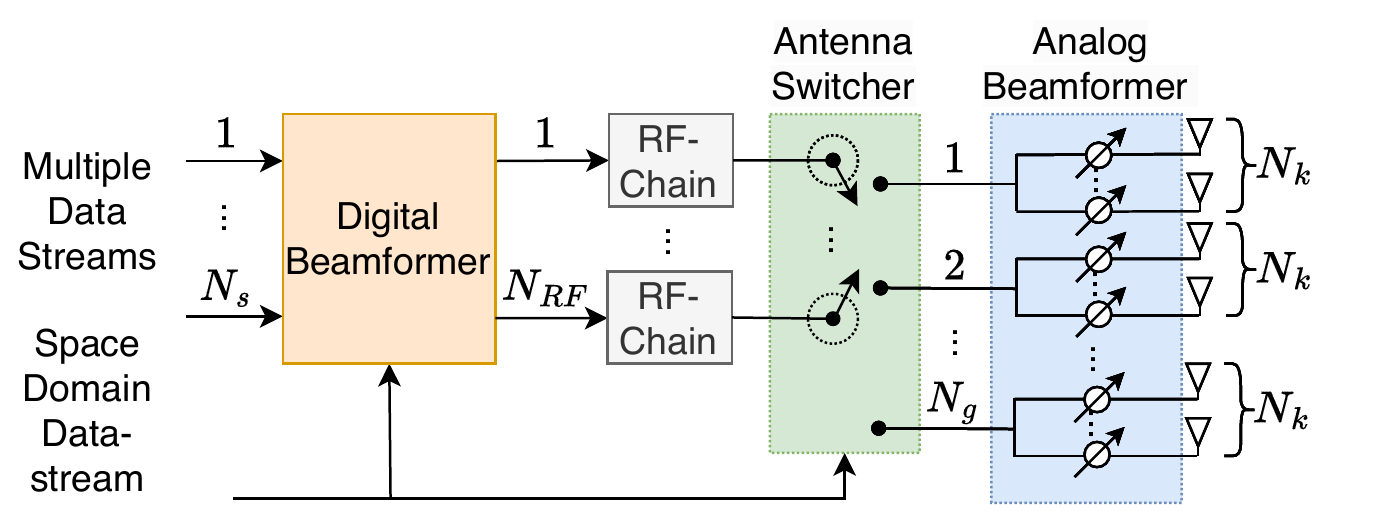}
  \caption{The architecture of GSM aided subarray hybrid beamforming.}
  \label{ImageSubarrayGSM}
\end{figure}

In this paper, a single user FDD mmWave MIMO system is considered with $N_t$ transmitting antennas at the BS and $N_r$ receiving antennas at the UE. As shown in Fig. \ref{ImageSubarrayGSM}, the $N_t$ transmitting antennas are equally split into $N_g$ groups with $N_k$ antennas in each group to align with the architecture of the GSM aided hybrid beamforming. Note that $N_g$ is larger than the number of RF chains $N_{RF}$.

The $N_s$ data streams are first processed by the digital beamformer to form $N_{RF}$ symbols as the input of RF chains. The outputs of RF chains are then mapped to $N_{RF}$ active antenna groups (AAGs) by the antenna switcher while the rest of the $(N_g-N_{RF})$ antenna groups remain silent. Notably, extra information can be carried by the choice of the AAGs, which is the key to the GSM. The total number of the legal AAGs selection schemes $M$ can be derived as follows \cite{lu2018low}.

\begin{equation} \label{EquationM}
  M = 2^{\left\lfloor \log_2\binom{N_g}{N_{RF}} \right\rfloor} \triangleq 2^{\left\lfloor \log_2\overbar{M} \right\rfloor}.
\end{equation}

We denote the AAGs selection scheme with an antenna connecting matrix $\mathbf{C}_m \in \{0,1\}^{N_t\times N_{RF}}, m\in\{1,\dots,M\}$. The element $(\mathbf{C}_m)_{ij}$ is set to one if and only if the $i^{\text{th}}$ antenna is activated and connected to the $j^{\text{th}}$ RF chain. In order to give a more intuitive idea, all possible $\mathbf{C}_m$ matrices are listed below for a vanilla example with $N_t=6, N_g=3, N_{RF}=2$.

\begin{equation}
  \mathbf{C}_1 =
	\left(
	\begin{matrix}
		1 & 0 \\
		1 & 0 \\
		0 & 1 \\
		0 & 1 \\
		0 & 0 \\
		0 & 0 \\
	\end{matrix}
	\right),\;\;
  \mathbf{C}_2 =
	\left(
	\begin{matrix}
		1 & 0 \\
		1 & 0 \\
		0 & 0 \\
		0 & 0 \\
		0 & 1 \\
		0 & 1 \\
	\end{matrix}
	\right),\;\;
  \mathbf{C}_3 =
	\left(
	\begin{matrix}
		0 & 0 \\
		0 & 0 \\
		1 & 0 \\
		1 & 0 \\
		0 & 1 \\
		0 & 1 \\
	\end{matrix}
	\right).
\end{equation}

It can be derived from equation (\ref{EquationM}) that $M$ is $2$ for the example above, which means two legal $\mathbf{C}_m$ matrices must be chosen from the three candidates. Such cases are common since $\overbar{M}=\binom{N_g}{N_{RF}}$ is most likely not a power of $2$. The maximum hamming distance is considered to choose the legal $\mathbf{C}_m$ following \cite{he2018spatial}.

\begin{equation} \label{EquationHamming}
  d_{pq} = \sum_{i=1}^{N_t}\left( \sum_{j=1}^{N_{RF}}\left(\mathbf{C}_p\right)_{ij} \oplus \sum_{j=1}^{N_{RF}}\left(\mathbf{C}_q\right)_{ij} \right)_, 1 \le p,q \le \overbar{M}.
\end{equation}

The hamming distance $d_{pq}$ between any two antenna connecting matrices is given by equation (\ref{EquationHamming}), where $(\cdot)_{ij}$ means the element at row $i$ and column $j$ and $\oplus$ represents the XOR operation. The legal $\mathbf{C}_m$ matrices are picked one by one, and the next matrix is always the one with the maximum average $d_{pq}$ from all the chosen matrices.

Further, each $\mathbf{C}_m$ is mapped with a different digital beamformer $\mathbf{D}_m \in \mathbb{C}^{N_{RF}\times N_s}$. Meanwhile, a constant modulus diagonal matrix $\mathbf{A} \in \mathbb{C}^{N_t \times N_t}$ is set as the global analog beamformer that consists of $N_t$ different phase shifters (PSs).

\begin{equation} \label{EquationAnalogBeamformer}
  \mathbf{A} = \frac{1}{\sqrt{N_k}}\text{diag}\left(e^{j\boldsymbol{\theta}_\mathbf{A}}\right),
\end{equation}
where $\boldsymbol{\theta}_\mathbf{A} \in \mathbb{R}^{N_t \times 1}$ controls the phase of all the $N_t$ PSs.

With the GSM aided hybrid beamformer architecture explained above, the downlink received signal at the UE can be derived as follows if the $m^{th}$ AAGs selection scheme is used.

\begin{equation} \label{EquationDownlinkTransmission}
  \mathbf{y} = \sqrt{P}\mathbf{H}\mathbf{A}\mathbf{C}_m\mathbf{D}_m\mathbf{s} + \mathbf{n},
\end{equation}
where $\mathbf{H} \in \mathbb{C}^{N_r \times N_t}$ is the downlink channel and $\mathbf{s} \in \mathbb{C}^{N_s \times 1}$ is the transmitting symbols. Besides, $P$ is the average power and $\mathbf{n} \sim \mathcal{CN}(0, \sigma^2\mathbf{I}_{N_r})$ is the additive white Gaussian noise.

Notably, the influential clustered Saleh-Valenzuela (SV) model is adopted for the mmWave MIMO channel as \cite{he2018spatial,lu2018low}.

\begin{equation} \label{EquationaSV}
  \mathbf{H} = \sqrt{\frac{N_tN_r}{N_{cl}N_{ray}}}\sum_{i=1}^{N_{cl}}\sum_{j=1}^{N_{ray}}\alpha_{i,j}\mathbf{a}_r(\theta^r_{i,j})\mathbf{a}_t(\theta^t_{i,j})^H,
\end{equation}
where $N_{cl}$ is the number of clusters and $N_{ray}$ is the number of rays in each cluster. $\alpha_{i,j}$ stands for the complex gain while the $\theta^r_{i,j}$ and the $\theta^t_{i,j}$ represent the angle of arrival (AoA) and the angle of departure (AoD), respectively. Besides, $\mathbf{a}_r(\theta^r_{i,j})$ and $\mathbf{a}_t(\theta^t_{i,j})$ are the normalized antenna response vector at the receiver and the transmitter. The following uniform linear array (ULA) model is adopted for both $\mathbf{a}_t(\theta^t_{i,j})$ and $\mathbf{a}_r(\theta^r_{i,j})$.

\begin{equation}
  \mathbf{a}(\theta) = \frac{1}{\sqrt{N}} \left[ 1, e^{j\frac{2\pi}{\lambda}\sin(\theta)}, \dots, e^{j\frac{2\pi}{\lambda}(N-1)\sin(\theta)} \right]^T,
\end{equation}
where $N$ is the number of transmitting or receiving antennas.

\section{The DL-based EFB Pipeline with GSM}  \label{Section-Algorithm}
In this section, the proposed GSM aided EFB pipeline with a jointly optimized network will be introduced and the loss function of the unsupervised training will be derived in detail.

\subsection{The Downlink Pilot Training under the GSM Architecture}
First of all, $L$ downlink pilots $\tilde{\mathbf{X}} \in \mathbb{C}^{N_t \times L}$ are sent by the BS for the channel estimation at the UE. Therefore, the downlink pilot transmission can be derived as follows.
\begin{equation} \label{EquationY}
  \tilde{\mathbf{Y}} = \mathbf{H}\tilde{\mathbf{X}} + \tilde{\mathbf{N}},
\end{equation}
where $\tilde{\mathbf{Y}} \in \mathbb{C}^{N_r\times L}$ is the received signal and $\tilde{\mathbf{N}}$ is the downlink additive white Gaussian noise. Note that $\tilde{\mathbf{X}}$ is designed via end-to-end training. The magnitude of $\tilde{\mathbf{X}}$ is set to $\sqrt{P/N_k}$ for simplicity while the phase of $\tilde{\mathbf{X}}$ is learned as the weight $\boldsymbol{\theta}_{\tilde{\mathbf{X}}} \in \mathbb{R}^{N_t \times L}$ of a fully connected (FC) layer without bias. Constrained by the GSM hybrid beamforming architecture, elements at positions with silent antenna are forced to zero so an additional mask $\mathbf{M}_p \in \{0,1\}^{N_t\times L}$ is necessary.

\begin{figure*}[!t]
  \centering
  \includegraphics[width=0.75\linewidth]{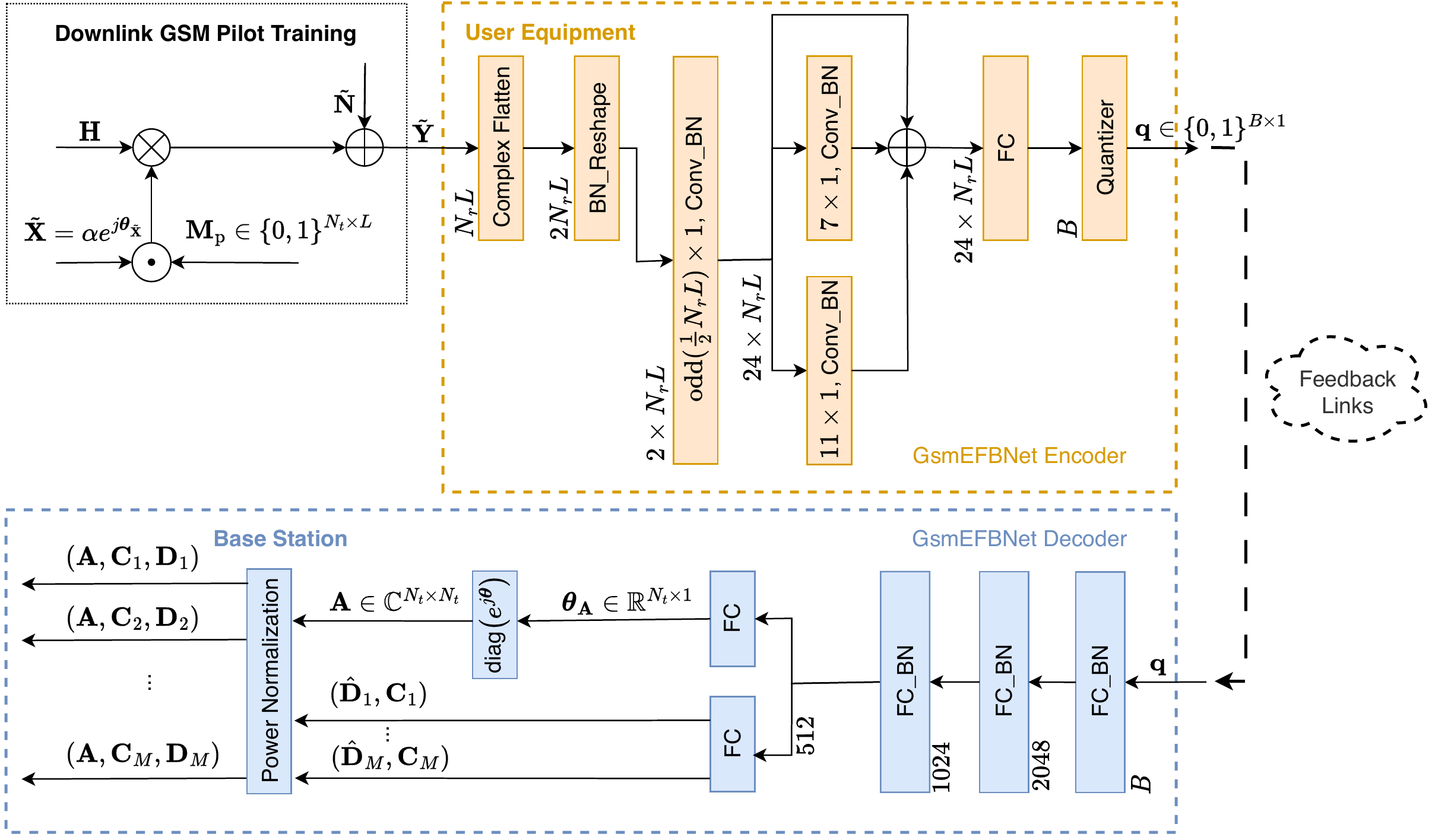}
  \caption{The proposed DL-based EFB pipeline with GSM hybrid beamforming. Note that the ReLU activation after each BN layer is left out for simplicity.}
  \label{ImageEndToEndSystem}
\end{figure*}

\begin{equation}
  \tilde{\mathbf{X}} = \sqrt{P/N_k}\left(\cos(\boldsymbol{\theta}_{\tilde{\mathbf{X}}}) + j\sin(\boldsymbol{\theta}_{\tilde{\mathbf{X}}})\right) \odot \mathbf{M}_p,
\end{equation}
where $\odot$ denotes the element-wise product. Remarkably, each column of $\mathbf{M}_p$ corresponds to the sum of columns of a specific $\mathbf{C}_m$. If the columns of $\mathbf{M}_p$ are badly chosen, all-zero rows are likely to occur and the channel information is partially lost since some antennas are inactive for all the pilots. In order to reduce such information loss, the columns of $\mathbf{M}_p$ are selected to maximize the hamming distance in equation (\ref{EquationHamming}).

\subsection{The Channel Estimation and Uplink Feedback}

As we can see from Fig. \ref{ImageEndToEndSystem}, the proposed \mbox{GsmEFBNet} encoder learns the uplink feedback bits $\mathbf{q}$ from the input $\tilde{\mathbf{Y}}$ at the UE and the channel estimation is done in an implicit way.

After the regular complex flattening, the input signals are sent into a convolutional expander with adaptive kernel size $\text{odd}(\frac{1}{2}N_rL)$, where $\text{odd}(x)$ gives the closest odd number to $x$. The channel is expanded from $2$ to
$24$ to enrich the feature. Two parallel branches with different convolutional kernel sizes $7$ and $11$ are applied for multi-resolution feature extraction. Note that such an adaptive multi-resolution network can extract channel information flexibly at different scales.

Finally, an FC layer is added to align the output dimension to the feedback capacity $B$ and a differential sign function is used as the quantizer to generate feedback bits following \cite{Sohrabi2021deep}.

\subsection{The Downlink GSM Aided Hybrid Beamforming}

As shown in Fig. \ref{ImageEndToEndSystem}, a pyramidal multilayer perceptron with intermediate output dimensions of $2048$, $1024$, and $512$ is used to construct channel information from the feedback bits $\mathbf{q}$. Based on the extracted channel feature, two FC layers are put in charge of learning the final hybrid beamformers.

Specifically, the phase vector $\boldsymbol{\theta}_{\mathbf{A}}$ is learned to generate the analog beamformer as equation (\ref{EquationAnalogBeamformer}). Further, $M$ independent matrices $[\hat{\mathbf{D}}_1, \dots, \hat{\mathbf{D}}_M]$ are learned so that each antenna connector $\mathbf{C}_m$ owns a specially designed digital beamformer. The following normalization is carried out for each digital beamformer to meet the power constraint $\left\Vert \mathbf{A}\mathbf{C}_m\mathbf{D}_m \right\Vert^2_F \le N_s$.

\begin{equation}
  \mathbf{D}_m = \frac{\sqrt{N_s}}{\left\Vert \mathbf{A}\mathbf{C}_m\hat{\mathbf{D}}_m \right\Vert_F} \hat{\mathbf{D}}_m.
\end{equation}

\subsection{The Unsupervised Loss for GSM Hybrid Beamforming}
In order to properly train the GsmEFBNet in an unsupervised way, the objective function of the GSM hybrid beamforming needs to be derived. Following \cite{lu2018low}, the beamforming achievable rate $R$ is defined as the mutual information (MI).

\begin{equation}
  R = I(\mathbf{y};\mathbf{s}|m) + I(\mathbf{y};m),
\end{equation}
where the $I(\mathbf{y};\mathbf{s}|m)$ is the amplitude-phase domain MI while the $I(\mathbf{y};m)$ is the spatial domain MI. Based on equation (\ref{EquationDownlinkTransmission}), the amplitude-phase domain MI can be derived as follows.

\begin{equation}
  I(\mathbf{y};\mathbf{s}|m) = \frac{1}{M} \sum_{m=1}^M\log_2\left(\left\vert \frac{\boldsymbol{\Sigma}_m}{\sigma^2} \right\vert\right),
\end{equation}
where the equivalent covariance matrix $\boldsymbol{\Sigma}_m$ is derived as:
\begin{equation}
  \boldsymbol{\Sigma}_m=\sigma^2\mathbf{I}_{N_r} + \frac{P}{N_s}\mathbf{H}\mathbf{A}\mathbf{C}_m\mathbf{D}_m\mathbf{D}_m^H\mathbf{C}_m^H\mathbf{A}^H\mathbf{H}^H.
\end{equation}

On the other hand, the spatial domain MI can be given as:
\begin{equation}
  I(\mathbf{y};m) = \frac{1}{M} \sum_{m=1}^{M}\int f(\mathbf{y}|m) \log_2 \left[\frac{f(\mathbf{y}|m)}{\frac{1}{M}\sum_{l=1}^{M}f(\mathbf{y}|l)}\right] \text{d}\mathbf{y}.
\end{equation}

The $I(\mathbf{y};m)$ above can be simulated via the Monte Carlo algorithm considering that $\mathbf{y} \sim \mathcal{CN}(0, \boldsymbol{\Sigma}_m)$ for the $f(\mathbf{y}|m)$. However, the lack of closed expression will block the gradient backpropagation. Therefore, the unsupervised loss is set as $L=-I(\mathbf{y};\mathbf{s}|m)$ instead of $L=-R$. Experiments show that the network can provide a satisfactory achievable rate $R$ even if the optimization focuses on the amplitude-phase domain MI.

\section{Results and Analysis} \label{Section-Experiment}

\subsection{Experimental Settings}

\begin{figure}[!t]
  \centering
  \includegraphics[width=\linewidth]{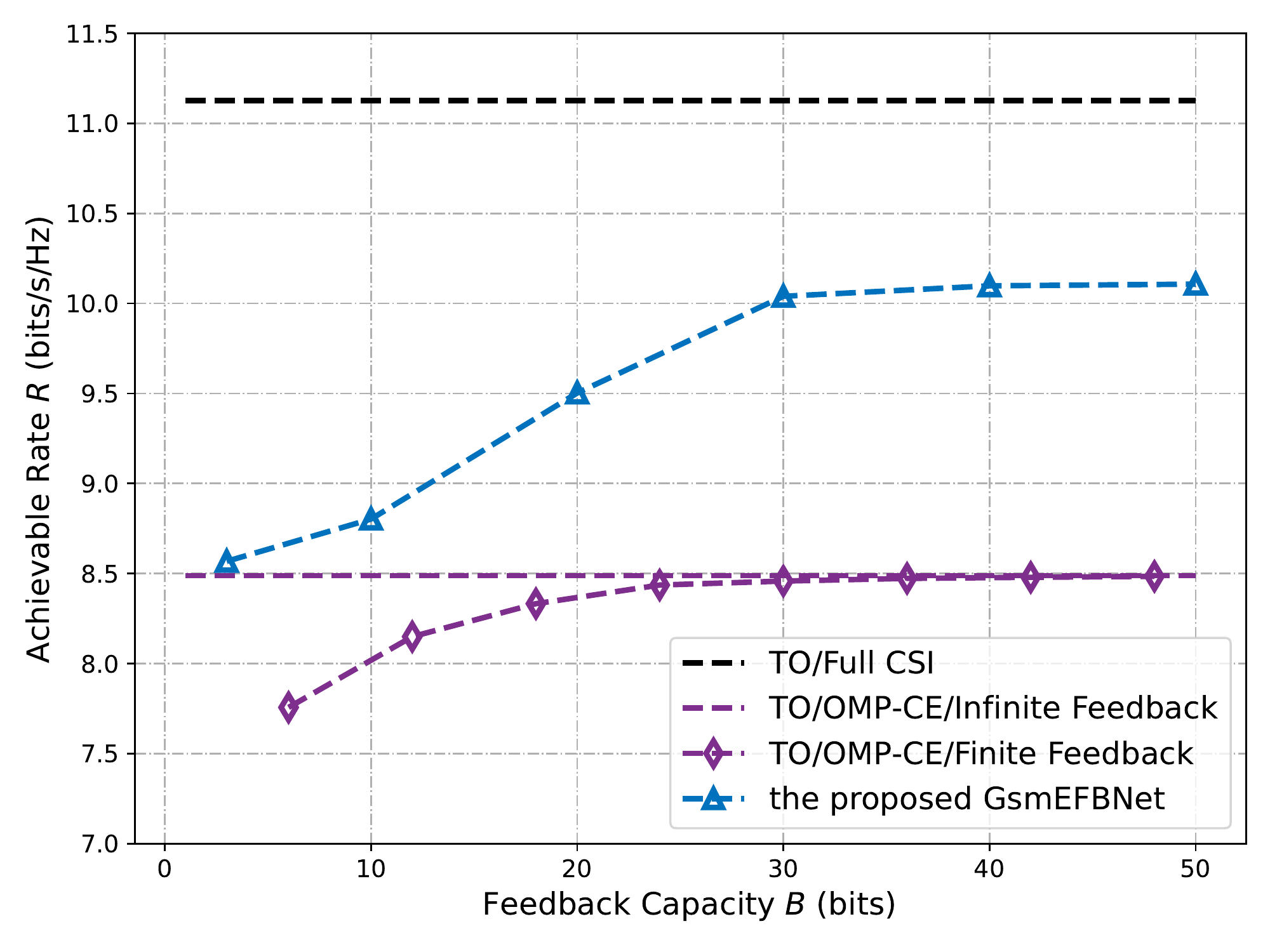}
  \caption{The achievable rates under different feedback capacities in a GSM aided FDD mmWave MIMO system with $N_t=16, N_r=4, N_s=N_{RF}=2, N_k=4$. The SNR is set to $10$dB for all the schemes.}
  \label{ImageResultOnB}
\end{figure}

The following simulation settings are given based on the SV model in equation (\ref{EquationaSV}) to generate the channel data. The number of clusters $N_{cl}$ and the number of rays in each cluster $N_{ray}$ are set to $2$ and $8$. The wavelength of the carrier is set to $5$mm and the azimuth angular spread is set to $7.5^\circ$. $[-30^\circ, 30^\circ]$ and $[-180^\circ, 180^\circ]$ are used as the range of azimuth sector angles for the transmitter and the receiver, respectively.

As for DL-related system settings, the Adam optimizer is adopted and the cosine annealing learning rate (LR) scheduler is applied with $10$ epochs of warmup. The initial LR is set to $5\times 10^{-4}$ while the minimum LR is set to $1\times 10^{-5}$. The batch size is set to $1000$ and the network is trained for $200$ epochs with $200$ batches in each epoch. In addition, $2000$ channel matrices are generated independently as the test dataset.

\subsection{Performance of the Proposed GsmEFBNet}

The following benchmarks are used to show the effectiveness of our proposed GsmEFBNet with fair comparisons.

\begin{itemize}
  \item [1)] \textit{TO/Full CSI}. The perfect CSI is given at the BS with ideal channel estimation and feedback, which is unrealistic. The turbo optimization (TO) algorithm proposed in \cite{lu2018low} is applied for conventional GSM aided hybrid beamforming.
  \item [2)] \textit{TO/OMP-CE/Infinite Feedback}. The downlink channel estimation with orthogonal matching pursuit (OMP) is considered at the UE. Ideal CSI feedback is assumed and the TO algorithm is used for the beamforming.
  \item [3)] \textit{TO/OMP-CE/Finite Feedback}. Compared with the second benchmark, the only difference is that practical limited feedback is considered with conventional quantization.
\end{itemize}

As shown in Fig. \ref{ImageResultOnB}, the proposed GsmEFBNet can outperform the conventional benchmark with OMP channel estimation and perfect feedback with even less than $10$ feedback bits when SNR is $10$dB. Moreover, the achievable rate of the GsmEFBNet with $B=40$ only drops for about $1$ bits/s/Hz compared with the conventional benchmark with perfect CSI. This shows that the channel learned by the adaptive multi-resolution encoder is of high quality.

\begin{figure}[!t]
  \centering
  \includegraphics[width=\linewidth]{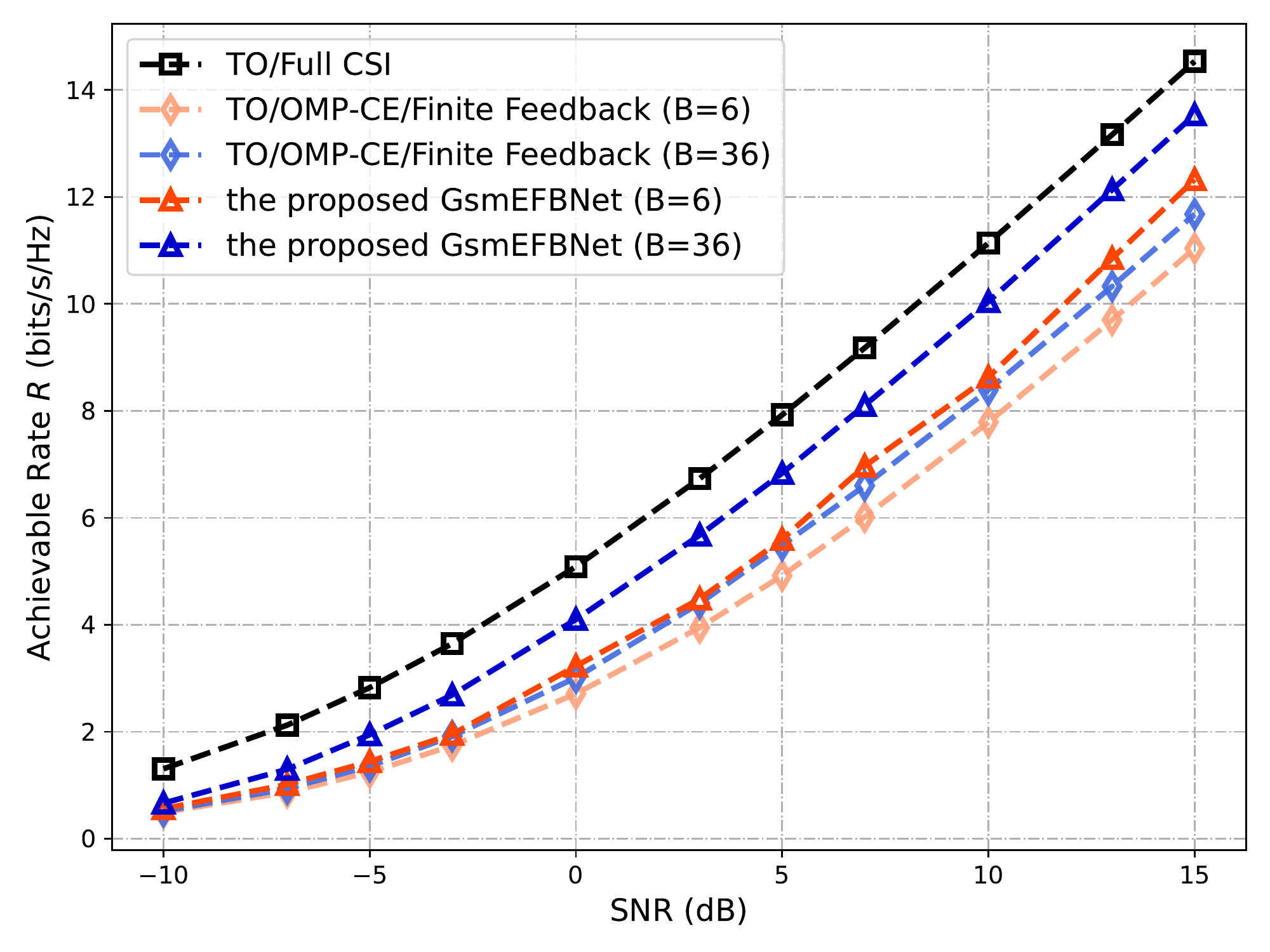}
  \caption{The achievable rates under different SNRs in a GSM aided FDD mmWave MIMO system with $N_t=16, N_r=4, N_s=N_{RF}=2, N_k=4$.}
  \label{ImageResultOnSNR}
\end{figure}

Further, the generalization capability for SNR is given in Fig. \ref{ImageResultOnSNR}. As we can see, the achievable rate performance of the proposed GsmEFBNet increases steadily as the SNR gets higher. The superiority of the proposed GsmEFBNet over the traditional benchmark under the same feedback capacity $B$ is very stable. Moreover, the achievable rate of GsmEFBNet with $B=6$ is even higher than the conventional benchmark with $B=36$. In a word, the robustness of the proposed GsmEFBNet under different SNRs is quite impressive.

\section{Conclusion} \label{Section-Conclusion}

In this paper, the DL-based channel estimation, feedback, and beamforming pipeline was first introduced to the GSM aided hybrid beamforming task for practical downlink CSI acquisition in FDD mmWave MIMO systems. A novel adaptive multi-resolution network named GsmEFBNet was specially designed to better extract CSI information from the masked pilots. Experiments showed that the proposed GsmEFBNet with limited feedback capacity could outperform the conventional benchmark with ideal feedback. Furthermore, the GsmEFBNet was robust to various SNRs as well.

\ifCLASSOPTIONcaptionsoff
  \newpage
\fi

\bibliographystyle{IEEEtran}
\bibliography{GsmEFBNet.bib}



\end{document}